\documentclass[twocolumn,showpacs,preprintnumbers,amsmath,amssymb]{revtex4}

\usepackage{graphicx}
\usepackage{dcolumn,subfigure,enumitem}
\usepackage{bm,color}
\usepackage{physics,natbib}

\begin{document}

\title{Chaos and Scrambling in Quantum Small Worlds}

\author{Jean-Gabriel Hartmann}
\author{Jeff Murugan}
\author{Jonathan P. Shock}

\affiliation{The Laboratory for Quantum Gravity \& Strings and, Department of Mathematics and Applied Mathematics, University of Cape Town,Private Bag, Rondebosch 7700, South Africa}
\date{\today}

\begin{abstract}
   Quantum small-worlds are quantum many-body systems that interpolate between completely 
ordered (nearest-neighbour, next-to-nearest-neighbour etc.) and completely random  interactions. 
As such, they furnish a novel new laboratory to study quantum systems transitioning between regular and 
chaotic behaviour. In this article, we introduce the idea of a quantum small-world network by starting from 
a well understood integrable system, a spin-$\frac{1}{2}$ Heisenberg chain. We then inject  
a small number of long-range interactions into the spin chain and study its ability to scramble quantum information
 using two primary devices: the out-of-time-order correlator (OTOC) and the spectral form factor (SFF). We find 
 that the system shows increasingly rapid scrambling as its interactions become progressively more random, with no 
 evidence of quantum chaos.
\end{abstract}


\maketitle

{\it Introduction}$-$Precipitated largely by studies of the gauge/gravity correspondence \cite{Maldacena:1997re}, the past five years have witnessed a tremendous global effort to understand the emergence of spacetime from the quantum properties of matter and information \cite{SciAm:2016}. This in turn has redefined the boundaries of contemporary condensed matter physics, high energy theory and even theoretical computer science, resulting in a number of remarkable discoveries. Among these are: (i) Maldacena, Shenker and Stanford's (MSS) bound on the growth rate of chaos in thermal quantum systems \cite{Maldacena:2015waa}, (ii) Kitaev's elaboration on the Sachdev-Ye spin-glass model \cite{Sachdev:1992fk} to the Sachdev-Ye-Kitaev (SYK) model \cite{Kitaev} of the quantum mechanics of $N$ Majorana fermions with infinitely long range disorder, and (iii) Witten's insight \cite{Witten:2016iux} that the perturbative structure of the large-$N$ limit of the SYK model is the same as colored random tensor models \cite{Gurau:2010ba}, thereby uncovering a new and tractable sweet-spot in between vector models and matrix models. While there are certainly many lessons to be drawn from these examples, two are of particular interest to us. The first, is that out-of-equilibrium interacting many-body systems appear to be generically chaotic and second, a new set of diagnostic tools are required to treat such systems\\

To date, the instrument of choice has been the commutator, $C_{\beta}(t)\equiv \langle[A(t),B(0)]^{2}\rangle_{\beta}$, or an equivalent OTOC, $F(t)$ related to the commutator through $C(t) = 2-2\mathrm{Re}(F)$. Here $A(t)$ and $B(t)$ are Heisenberg operators and $\langle\ldots\rangle_{\beta}$ denotes the thermal average performed at temperature $T = 1/k_{B}\beta$. Originally introduced in the context of superconductivity in 1969 \cite{Larkin:1969aa},  the OTOC has recently been repurposed in the context of chaotic many-body systems where $C_{T}(t)\sim \hbar^{2}e^{2\lambda t}$ with $\lambda$ identified as the {\it quantum Lyapunov exponent}. This was instrumental in establishing the MSS bound $\lambda\leq 2\pi k_{\mathrm{B}}T/\hbar$. However unlike their more familiar time-ordered counterparts, OTOCs are less well understood and much more subtle \cite{Hashimoto:2017oit}. To understand better what information they do, and do not, encode, it is imperative that they be scrutinised  more broadly. With this in mind, we would like to address the question: {\it What can we learn from, and about, OTOCs  in an interacting many-body system that is somewhere between completely regular and completely disordered?}\\

The first step in answering this question is to set up a system that interpolates between these two extremes, with some control parameter. One useful way to treat interactions in a dynamical system is to take the individual components of the system (electrons in an atomic sample, computer servers in the internet or even people in a social network, etc.) to be nodes, $v_{i}$, connected by edges, $e_{ij} = \{v_{i},v_{j}\}$, that encode their interactions. Together, the collection of nodes $V$ and edges $E\subseteq V\otimes V$ form a {\it graph}, $G(V,E)$, or when referring to `real world' entities, a {\it network} \cite{Estrada:2013aa}. In this language \cite{ESSAM:1970zz}, the details of the individual interactions among components are secondary to its global features, like the  topology of the network. Such features are of obvious importance when the dynamical system in question is a quantum many-body problem, essentially because here they are intimately tied to questions of thermalisation, entanglement and the spread of information in the system.\\

In the set of possible graphs with $n$ nodes and $m$ edges, we distinguish two classes: the nodes in {\it k-regular graphs} each have degree $k$ giving a total of $m=nk/2$ edges in the graph. Such regular lattices are characterised by a high degree of localised clustering. This is to be contrasted with {\it random (Erd\"os-Renyi) graphs} in which each pair of nodes is assigned an edge with some probability, and which display characteristically small path lengths and correspondingly efficient information spreading throughout the network. In this language, periodic quantum spin-chains with nearest neighbour interactions are 2-regular closed graphs, while the SYK model  is an example of an edge-weighted {\it complete} random graph whose edges are assigned with a Gaussian probability.\\


{\it Classical small worlds$-$}With the goal of studying the onset of chaos in a controlled environment, we construct interacting quantum lattice systems which can be tuned toward, or away from, integrability. At the extremes of this family of models are (i) regular, nearest-neighbour, lattices, and (ii) completely connected lattices, both of which have some aspect of regularity about them. Additionally, we would like to add randomness into the lattice structure in a controlled way.\\

Our construction will be based on the small-world model defined by Watts and Strogatz in their seminal work \cite{Watts1998}.  Starting with an $N$-site lattice (nodes in the graph) and $k/2$-nearest-neighbour interactions \footnote{Here $k$ is the number of edges attached to each node.} (edges) we proceed by going to each node, $n_i$, in turn, and iterating through each edge connected to that node. With probability $p$, a given edge in the graph is disconnected, and is then replaced with an edge going from $n_{i}$, to a random node, $n_j$ ($i\ne j$) in the graph. This leaves the total number of edges in the network unchanged, but creates random long-range connections which gives the graph its `small-world' property. Examples of this procedure for different values of $k$ and $p$ are shown below.\\

\begin{center}
	\includegraphics[width=5cm]{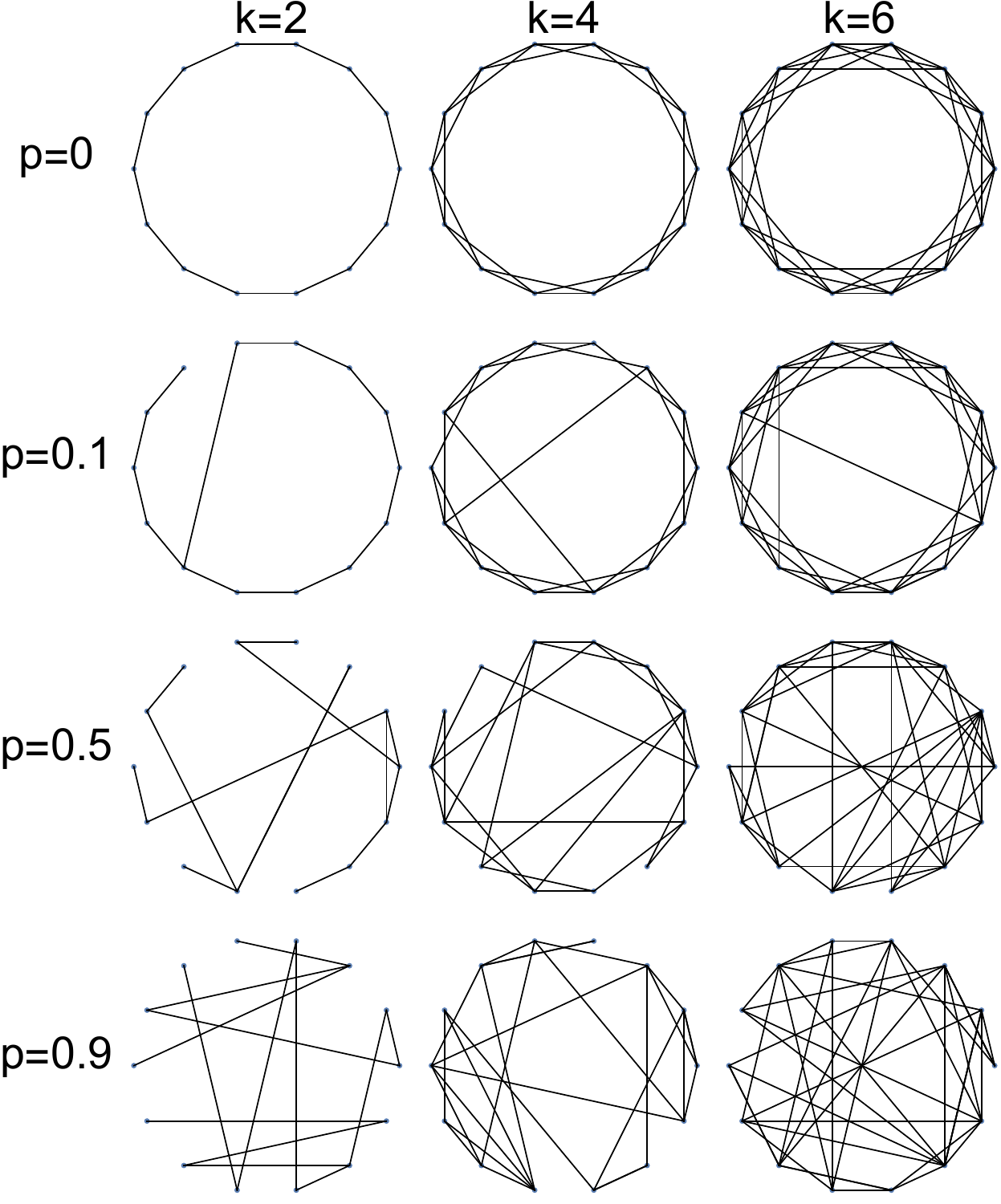}
\end{center}

As long as $p$ is non-zero, we will consider the graph to be small-world, although when $p\ll1/N$ the likelihood that the network remains in its original state is high. For this reason, $p$ is not an appropriate parameter to use to characterise how far the network is from a regular, integrable, system. For this purpose, we will discuss two metrics of small-worldness, both of which depend on the clustering coefficient and the path length in the graph. The clustering coefficient, $C$ is defined as the average of $C_i\equiv2E_i/(k_i(k_i-1))$ over all nodes $i$, where $E_i$ is the number of edges between the direct neighbours of node $i$, and $k_i$ is the total number of edges attached to node $i$. This  measures the average connectedness of a node to its neighbours in proportion to its overall connections across the entire network. The path length of the graph, $L\equiv\sum_{i\ne j} d_{ij}/(N(N-1))$ is defined as the network average of the shortest geodesic distance between any two nodes in the graph. On average, both $C$ and $L$ decrease with $p$, though the average path length decreases much more rapidly for small $p$ compared to the clustering coefficient. A small-world network is therefore sometimes considered as a network with small $L/L_0$ but relatively large $C/C_0$ where $L_0$ and $C_0$ are the values of $L$ and $C$ for the regular graphs with $p=0$.\\ 

\begin{figure}
	\begin{center}
		\includegraphics[width=4cm]{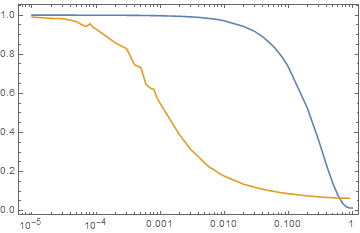}
		\caption{A plot of the mean shortest path length, $L(p)/L(0)$, and the mean clustering coefficient, $C(p)/C(0)$, as functions of the re-wiring probability, $p$. The $L(0)$ and $C(0)$ terms act as normalisation factors with respect to the regular, $p=0$, graph. Calculations were performed on a large, 5-regular, graph of 1000 vertices, averaged over 100 instantiations.}
	\end{center}
\end{figure}

{\it Quantum small-worlds$-$}Thus far, the small-world networks we have described are classical. Let's now define a quantum analog. In order to investigate the effect of network topology on the dynamics of a quantum many-body system, we consider a network of spin-$\frac{1}{2}$ particles at each vertex, with edges representing spin-exchange interactions between the respective lattice positions. To this end, consider the Hamiltonian,
\begin{equation}\label{ham}
H = -\sum_{i=1}^{N}\sum_{j = i+1}^{N}\sum_{k = 1}^{3}A_{ij} S_i^k S_j^k \; ,
\end{equation}
with $N$ the number of vertices in the network, $A_{ij}$ the $ij$-th element of the graph adjacency matrix, and $S_i^k = \frac{1}{2}\sigma_i^k$ the $k$'th Pauli spin-$\frac{1}{2}$ matrix acting at vertex $i$. Correspondingly, $H \subset GL(2^N \times 2^N,\mathbb{C})$. The network topology is encoded in the adjacency matrix. In particular, if we fix the coupling strengths to be uniform throughout the spin network, and normalize to unity, then $A_{ij}$ is an $N\times N$ matrix populated by either 1's or 0's. By turning on appropriate matrix elements, we can tune the spin network through various topologies. Figure 2. displays various regular network configurations with the corresponding adjacency matrices and the numerically computed Hamiltonian eigenvalue spectrum, in full agreement with known results. Now let's probe the system as we inject some (small) number of long range interactions between lattice sites. Specifically, we would like to study how the information of a kick given to one of the spins at some initial site is scrambled as we tune the system from regular (and integrable) through random (and chaotic). To do so, we implement the Watts-Strogatz protocol outlined in section II on the 1-dimensional lattice. Figure 3. displays results for an 11-site lattice with fixed next-to-nearest neighbour coupling. We numerically diagonalize the rewired Hamiltonian and compute its eigenvalue spectrum. The regular lattices of Figure 2. all correspond to re-wiring probability $p=0$. Note also that the spectrum, even for large values of $p$ is nearly identical to the regular chain. This behaviour is also observed at larger values of $k$. \\

\begin{figure}
\begin{center}
   \includegraphics[width=6cm]{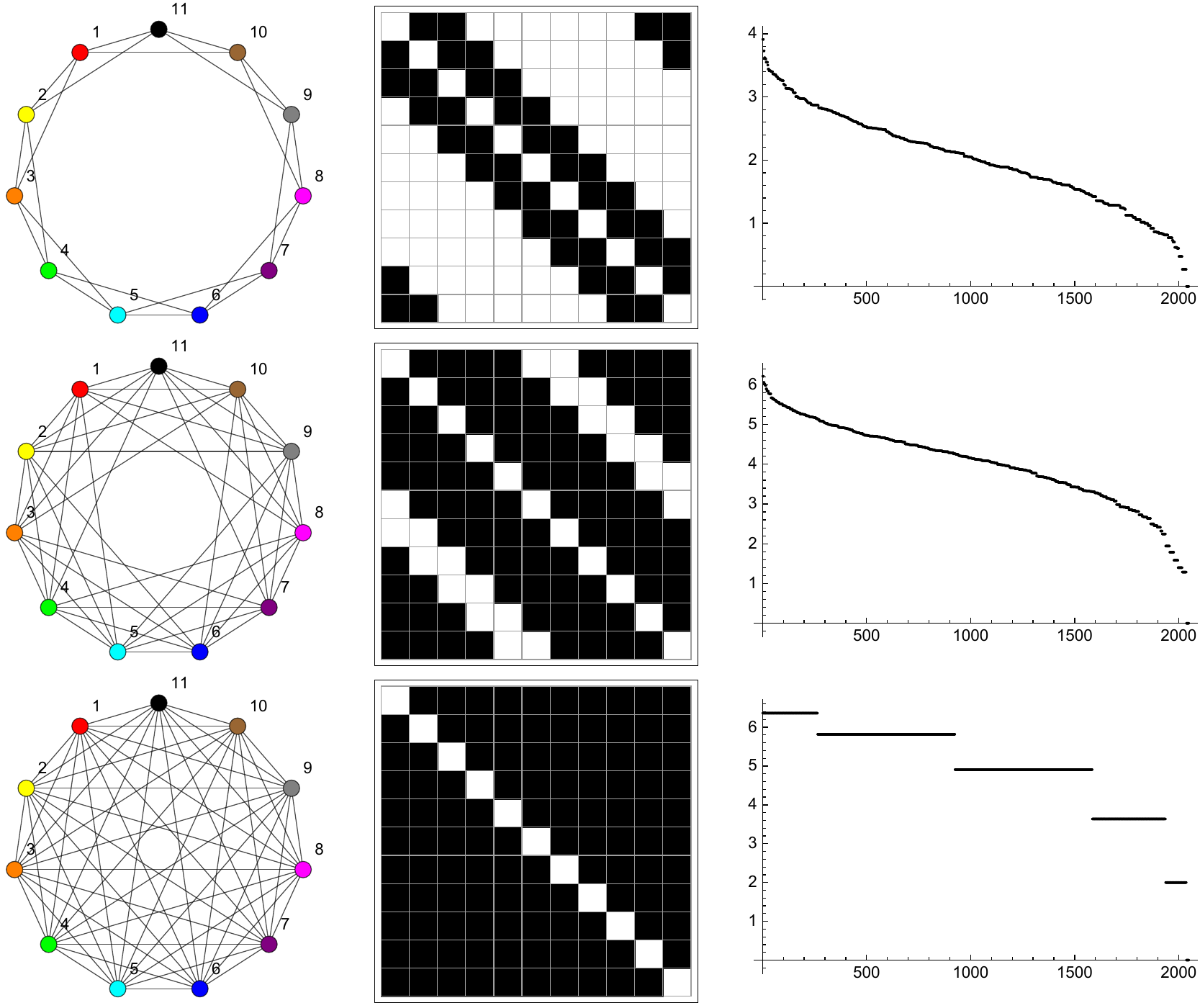}
   \caption{From left to right: The network diagram with each edge representing a spin-spin coupling; adjacency matrix, $A_{ij}$ with black squares corresponding to edges between nodes; Hamiltonian occupancy and eigenvalue spectrum. All chains have $N=11$ sites and differ only in the range of the spin-spin interaction. From top to bottom, $(k,C,L) = (4,0.5,1.8),(8,0.75,1.2)$ and $(10,1.0,1.0)$.}
\end{center}
\end{figure}

\begin{figure}
\begin{center}
   \includegraphics[width=6cm]{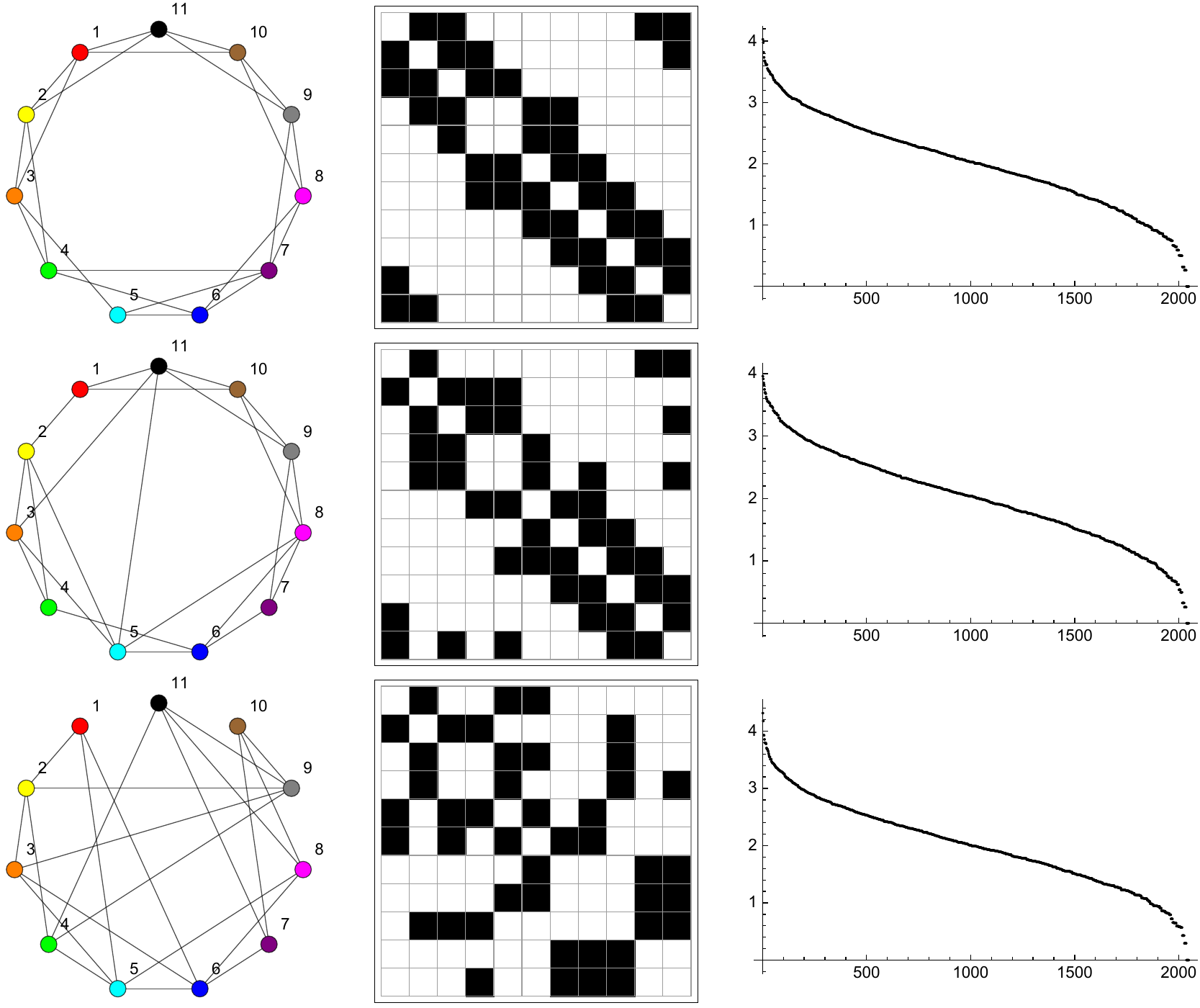}
   \caption{Implementing the Watts-Strogatz protocol. From left to right: The network diagram, adjacency matrix, Hamiltonian occupancy and eigenvalue spectrum. All chains have $N=11$ sites with spin-spin interaction range fixed at $k=4$ and differ only in the re-wiring probability, $p$. From top to bottom, $(p,C,L) = (0.1,0.45,1.76), (0.5,0.41,1.73)$ and $(0.75,0.23,1.67)$}
\end{center}
\end{figure}
Scrambling, the tendency of a many body quantum system to delocalize quantum information over all its degrees of freedom, is diagnosed by $C(t)$, the thermally averaged square of a commutator or, alternatively, the OTOC, $F(t) \equiv \langle A^{\dagger}(t)B^{\dagger}(0)A(t)B(0)\rangle$ for some choice of unitary Heisenberg operators $A(t)$ and $B(t)$ in the system. To begin our study of scrambling in quantum small-world networks, we will compute the infinite temperature 
four-point OTOC,
\begin{equation}\label{otoc}
	C_{\beta = 0}(t) = \ev{S_i^z(0)S_j^z(t)S_i^z(0)S_j^z(t)}{\psi}_{\beta = 0}\,.
\end{equation}
Employing the notion of quantum typicality, the expectation value is approximated \cite{steinigeweg_spin-current_2014} by the overlap of two time-evolved states, $S_i^z(0)S_j^z(t)\ket{\psi}$ and $ S_j^z(t)S_i^z(0)\ket{\psi}$, where $\ket{\psi}$ is a random pure state drawn from the $2^N$-dimensional Hilbert space. Here $S_i^z(0) = S_i^3$ is the un-evolved spin operator defined above, and $S_j^z(t) = e^{iHt}S_j^z(0)e^{-iHt}$ the time-evolved Heisenberg spin operator. Our numerical results for the computation of the OTOC for the small-world chain and for various values of the re-wiring probability are summarized in Figure 4. 
\begin{figure}
\begin{center}
	\includegraphics[width=7cm]{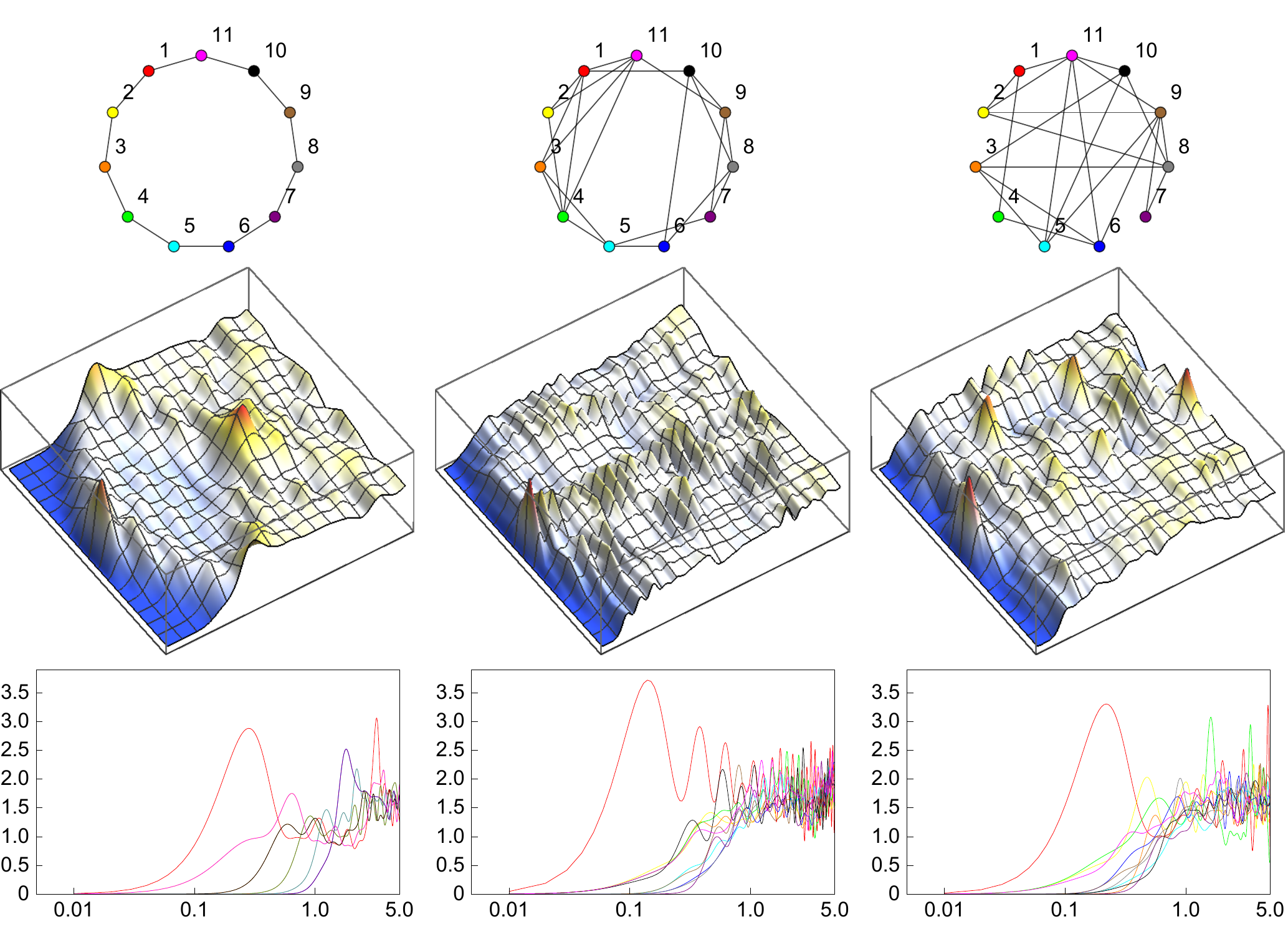}
	\caption{Numerical results for the OTOC, $C_{ij}(t) = 2(1 - Re(C_0(t)))$ as a function of time and vertex position. All computations were performed on an 11-site lattice with varying number of random re-wirings of the lattice following the Watts-Strogatz protocol. Middle row from left to right:  $C_{ij}(t)$ for lattice configurations (vertex degree $k$, re-wiring probability $p$) $= (2,0), (4,0.25), (4,0.75)$. The left most configuration is a closed XXX Heisenberg spin-chain with nearest neighbour interactions. In the bottom row we display vertex-wise correlators $C_{1j}(t)$ for an initial disturbance at site 1. In all cases, at early times $C_{1j}\sim t^{b}$ where $1.76\leq b\leq 6.23$, $1.76\leq b\leq 3.22$ and $1.73\leq b\leq 3.42$ for $p=0,0.25$ and $0.75$ respectively. }
\end{center}
\end{figure}

Previous studies of many-body integrable to chaotic transitions \cite{Garcia-Garcia:2017bkg}, for example, in the  the (mass-deformed) SYK model uncovered a tension between the OTOC and typical random matrix theory (RMT) diagnostics. In part, this is a reflection of the nature of the two sets of tools; the OTOC captures early time, quantum mechanical features of the model whereas RMT captures late time, statistical features. To reconcile these two observations, in the context of black hole information scrambling, the authors of \cite{Cotler2017} proposed an alternative diagnostic in the {\it spectral form factor} (SFF). As the analytical continuation of the thermal partition function the SFF, $g(t,\beta)$, has two desirable properties: (i) at late times it displays RMT behaviour and (ii) because it has a quantum mechanical flavor, it is closer to the OTOC description of quantum chaos than standard RMT measures. Concretely, we compute the {\it annealed} SFF \cite{Cotler2017},
\begin{equation}\label{sff}
	g(t;\beta) = \frac{\ev*{\abs{Z(\beta, t)}^2}_J}{\ev{Z(\beta)}^2_J}
\end{equation}
where $\ev{\cdot}_J$ is the disorder-averaged expectation value. In the infinite temperature ($\beta \to 0$) limit, this expression reduces to $g(t;0) = \ev*{\abs{Z(0, t)}^2}_J$ with 
\begin{equation}
	\abs{Z(0, t)}^2 = \sum_{m=1}^{2^N} \sum_{n=1}^{2^N} e^{i(E_m-E_n)t}
\end{equation}
as the magnitude of the analytically-continued partition function, $Z(\beta)$. The results of our numerical computation for $g(t,0)$ for various re-wirings of the network according to the Watts-Strogatz protocol are plotted in Figure 5.\\

\begin{figure}
	\begin{center}
		\includegraphics[width=9cm]{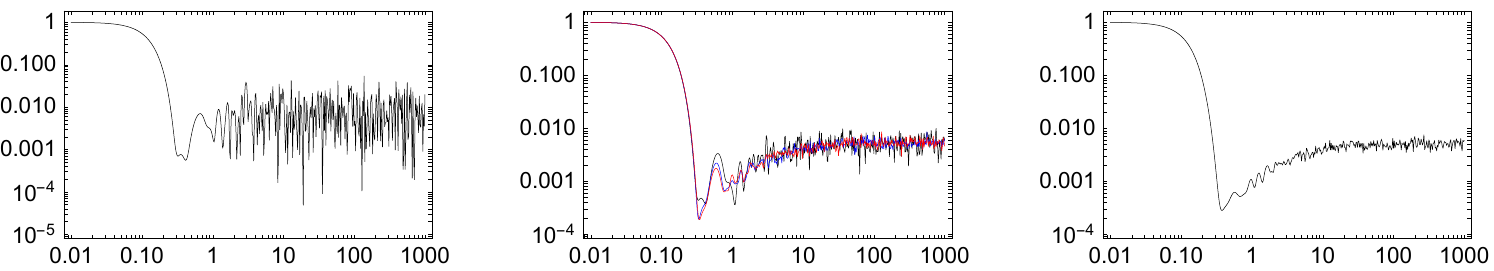}
			\caption{The infinite temperature spectral form factor $g(t,\beta=0)$ for various re-wirings 
			of the network. On the left is the SFF for zero re-wirings. The SFF does not show any dip or 
			ramp regimes. The middle plot shows the SFF for 1, 2 ({\color{blue}{blue}}) and 3 
			({\color{red}{red}}) re-wirings. The dip-linear ramp-plateau behaviour is immediate and more 
			pronounced for increasing randomness. On the far right, the SFF computed for 20 re-wirings 
			of the network where $p\sim 1$ clearly displays a linear ramp connecting the dip to the plateau 
			at $t\sim 10$. We have checked also that the onset of the plateau remains at $t\sim 10$ for 
			$N=6,7,8$ and $9$.}
	\end{center}
\end{figure}

{\it Discussion$-$}Scrambling in quantum systems appears to be the result of a confluence of a number of properties of the system: randomness, interactions, disorder and chaos. Disambiguating between these is of paramount importance to the understanding to the physics of a number of important problems, from black holes to quantum computing. This article details our study of a quantum small world - the quantum analog of the famed Watts-Strogatz model - which we introduce to understand aspects of the transition of an integrable system into (quantum) chaos. In this first salvo, we have focused our attention on setting up the model and carry out some numerical experiments with several chaos diagnostics. The model itself consists of a 1-dimensional $k$-local, $N$-site spin chain into which is inserted, for fixed $k$, a small number of long range interactions with some probability $p$. For $p=0$ and small values of $k\geq 4$ the spin chain is highly cliquey and localises interactions to neighbourhoods, as is inferred by the proliferation of triangles in the top left corner of Figure 2.  The infinite temperature OTOC for both the integrable nearest neighbour (see the first column of Figure 4.) and nonintegrable next-to-nearest re-wiring interacting chains converges on $C_{ij}=2$ (or $C_{0}=0$) at late times and for generic initial states. We have checked also that convergence happens faster with increasing $k$. This signals that the system does indeed scramble {\it without chaos}, independently confirming the results reported in \cite{Iyoda:2017pxe} obtained using numerical exact diagonalization.\\ 

 Next, we turn on some number of long-range couplings by a random re-wiring of the network edges with $p\neq0$, and computed the associated OTOC \eqref{otoc}. In each case, we find that the early time behaviour of the OTOC is polynomial in $t$. As we increase the small-worldness of the model, the OTOC converges increasingly rapidly on $C_{ij}(t)=2$ at late times. Correspondingly, the system rapidly delocalises an initial kick at vertex 1, again without signs of chaos. To check this conclusion,  we then computed the spectral form factor $g(t,\beta)$ that is supposed to interpolate between the early time OTOC behaviour and late time random matrix theory behaviour of a genuine quantum chaotic system. Having computed the SFF for a number of re-wirings of the spin chain (corresponding to increasing randomness) we found that it displayed the dip-linear ramp-plateau behaviour charateristic of quantum chaos. However, unlike in a truly chaotic system, the onset of the plateau (or eqivalently the length of the linear ramp regime) does not scale with $N$ \footnote{We are grateful to Dario Rosa for a discussion of this point.}, and so this behaviour is more reminiscent of the random but integrable SYK$_{2}$ model found in \cite{Lau:2018kpa}.\\

Some further comments are in order. Firstly, the OTOC and its analytic properties are best understood in the large-$N$ (and in the SYK model, simultaneously the large $q$) limit. Since our study here is restricted to $k\leq N$ and $N\leq 11$, strictly speaking, we have neither. What we have is a few-body sparse quantum system. In such systems, even though turning on $p$  drastically changes the properties of the system, further variation of $0<p\leq1$ has relatively little effect. Evidently, the re-wiring probability is not a good parameterization of small-worldness for small values of $N$. Fortunately, the clustering coefficient and mean path length on the network allow for the construction of alternative parameterizations. In particular, denoting by $C_{rand}$ and $L_{rand}$ the values of $C$ and $L$ for a completely random graph with the same number of nodes and edges as in the small-world construction, we can define $\gamma\equiv C/C_{rand}$ and $\lambda\equiv L/L_{rand}$. Their ratio $\sigma\equiv \gamma/\lambda$ furnishes another measure of ``small-worldness", in the sense that a small-world network is characterised by $C\gg C_{rand}$ and $L\sim L_{rand}$, leading to $\sigma>1$. Yet another measure of small-worldness can be defined as $\omega\equiv L_{rand}/L - C/C_{latt}$, with $C_{latt}$ the clustering coefficient for a regular lattice. This has the advantage of being a monotonically increasing function of $p$ and should be contrasted with $\sigma$ which peaks at the point where the system is optimally small-world and then decreases again as we tune towards a random lattice with many long-range connections. Either way, understanding how the chaos diagnostics vary with these parameters would be an important refinement of our conclusions.\\

Second, our quantum model confirms our intuition inherited from the classical Watts-Strogatz model, namely that the introduction of a small number of long range interactions into the system rapidly delocalizes information in the network. However, quantum mechanics is much more subtle than classical systems. For example, most of our results hold strictly in the infinite temperature limit where the computations simplify dramatically. These simplifications are lost at finite temperatures and our conclusions need to be explicitly checked in this regime. Finally, while the model we study here is clearly a toy one, it is worth pointing out that such systems are not too far from realizable in recent table-top cold-atom experiments  with cavity QED \cite{Swingle:2016var}. It would be very exciting to be able to physically test this protocol in the near future.\\

 In any event, we have only just scratched the surface of these models, and that, with a small toothpick. Some of these questions we will return to in a forthcoming article \cite{Hartman2}, but it goes without saying that there remains much more to be done.\\


{\it Acknowledgements$-$}We would like to thank Micha Berkooz, Tim Gebbie, Chen-Te Ma, Javier Magan, Dario Rosa, Joan Simon and Masaki Tezuka for very useful discussions. JGH is supported by a graduate fellowship from the National Institute for Theoretical Physics. JM is supported by the NRF of South Africa under grant CSUR 114599 and the National Science Foundation under Grant No. NSF PHY-1748958. JM would like to thank the organisers and participants of the “Chaos and Order 2018” program at the KITP of the University of California, Santa Barbara for a stimulating and productive environment during the final stages of this work.


\begin{thebibliography}{9999}

\bibitem{Maldacena:1997re} 
  J.~M.~Maldacena,
  Int.\ J.\ Theor.\ Phys.\  {\bf 38}, 1113 (1999)
  [Adv.\ Theor.\ Math.\ Phys.\  {\bf 2}, 231 (1998)]
  doi:10.1023/A:1026654312961, 10.4310/ATMP.1998.v2.n2.a1
  [hep-th/9711200].

\bibitem{SciAm:2016}
https://www.scientificamerican.com/article/tangled-up-in-spacetime/

\bibitem{Maldacena:2015waa} 
  J.~Maldacena, S.~H.~Shenker and D.~Stanford,
  JHEP {\bf 1608}, 106 (2016)
  doi:10.1007/JHEP08(2016)106
  [arXiv:1503.01409 [hep-th]].

\bibitem{Sachdev:1992fk} 
  S.~Sachdev and J.~Ye,
  Phys.\ Rev.\ Lett.\  {\bf 70}, 3339 (1993)
  doi:10.1103/PhysRevLett.70.3339
  [cond-mat/9212030].

\bibitem{Kitaev}
A. Kitaev, “A simple model of quantum holography.” http://online.kitp.ucsb.edu/online/entangled15/kitaev/,http: //online.kitp.ucsb.edu/online/entangled15/kitaev2/. Talks at KITP, April 7, 2015 and May 27, 2015.

\bibitem{Witten:2016iux} 
  E.~Witten,
  arXiv:1610.09758 [hep-th].
  
\bibitem{Gurau:2010ba} 
  R.~Gurau,
  Annales Henri Poincare {\bf 12}, 829 (2011)
  doi:10.1007/s00023-011-0101-8
  [arXiv:1011.2726 [gr-qc]].
 
 \bibitem{Larkin:1969aa}
 A. I. Larkin and Y. N. Ovchinnikov, JETP 28, 6 (1969): 1200-1205.
  
\bibitem{Hashimoto:2017oit} 
  K.~Hashimoto, K.~Murata and R.~Yoshii,
  JHEP {\bf 1710}, 138 (2017)
  doi:10.1007/JHEP10(2017)138
  [arXiv:1703.09435 [hep-th]].
   
\bibitem{Estrada:2013aa}
E. Estrada,
[ arXiv:1302.4378v2]  

\bibitem{ESSAM:1970zz} 
  J.~W.~Essam and M.~E.~Fisher,
  Rev.\ Mod.\ Phys.\  {\bf 42}, 271 (1970).
  doi:10.1103/RevModPhys.42.271
 
 \bibitem{Watts1998}
 D. J. Watts and S. H. Strogatz.  
 Nature,393,440 (1998)
 
\bibitem{steinigeweg_spin-current_2014}
R.~Steinigeweg, J.~Gemmer and W.~Brenig.
Phys.\ Rev.\ Lett.\  {\bf 112}, 120601 (2014)
doi:10.1103/PhysRevLett.112.120601
\newblock Spin-current autocorrelations from single pure-state propagation.
[arXiv:1312.5319v2 [cond-mat.str-el]].

\bibitem{Garcia-Garcia:2017bkg} 
  A.~M.~García-García, B.~Loureiro, A.~Romero-Bermúdez and M.~Tezuka,
  Phys.\ Rev.\ Lett.\  {\bf 120}, no. 24, 241603 (2018)
  doi:10.1103/PhysRevLett.120.241603
  [arXiv:1707.02197 [hep-th]].
  
\bibitem{Cotler2017}
J.S.~Cotler, G.~Gur-Ari, M.~Hanada, J.~Polchinski, P.~Saad,
S.H.~Shenker, D.~Stanford, A.~Streicher, and M.~Tezuka.
 JHEP {\bf 2017}, 118 (2017)
 doi:10.1007/JHEP05(2017)118 
\newblock Black Holes and Random Matrices.
[arXiv:1611.04650v3 [hep-th]].
  
\bibitem{Swingle:2016var} 
  B.~Swingle, G.~Bentsen, M.~Schleier-Smith and P.~Hayden,
  Phys.\ Rev.\ A {\bf 94}, no. 4, 040302 (2016)
  doi:10.1103/PhysRevA.94.040302
  [arXiv:1602.06271 [quant-ph]].

\bibitem{Iyoda:2017pxe} 
  E.~Iyoda and T.~Sagawa,
  Phys.\ Rev.\ A {\bf 97}, no. 4, 042330 (2018)
  doi:10.1103/PhysRevA.97.042330
  [arXiv:1704.04850 [cond-mat.stat-mech]].
  
\bibitem{Lau:2018kpa} 
  P.~H.~C.~Lau, C.~T.~Ma, J.~Murugan and M.~Tezuka,
  arXiv:1812.04770 [hep-th].
  
\bibitem{Hartman2}
   J-G. Hartmann, J. Murugan and J.P. Shock,
   ``More on scrambling, randomness and chaos in quantum small worlds,"
   In preparation, 2019   
   
\end{thebibliography}
\end{document}